\documentclass[a4paper,11pt]{article}
\usepackage[utf8]{inputenc}

\usepackage{geometry}
\usepackage{xcolor}
\usepackage{amsmath, array, amssymb, amsfonts,amsthm}
\usepackage{xfrac}
\usepackage[inline]{enumitem}

\usepackage[french, english]{babel}
\usepackage[all]{xy}

\usepackage{txfonts}  

\usepackage{sectsty}
\sectionfont{\centering}
\subsectionfont{\centering}
\subsubsectionfont{\centering}

\usepackage{booktabs}
\usepackage{caption}
\usepackage{dsfont}
\usepackage{mathtools}
\usepackage{slashed}

\usepackage[hidelinks]{hyperref}

\usepackage{textcomp}

\usepackage{multicol}
\usepackage[square,numbers,sort,compress,semicolon,merge]{natbib}
\let\cite\citep 
\usepackage{hyperref}
\hypersetup{colorlinks=true, urlcolor=blue, citecolor=blue, linktoc=page}

\makeatletter
\renewcommand*\env@matrix[1][\arraystretch]{%
  \edef\arraystretch{#1}%
  \hskip -\arraycolsep
  \let\@ifnextchar\new@ifnextchar
  \array{*\c@MaxMatrixCols c}}
\makeatother

\renewcommand\P{\mathcal{P}}
\newcommand\M{\mathcal{M}}

\newcommand\RR{\mathbb{R}}
\newcommand\CC{\mathbb{C}}
\renewcommand\1{\textbf{1}}

\renewcommand\H{\mathcal{H}}

\renewcommand\S{\mathcal{S}}

\newcommand\SU{\mathcal{SU}}
\newcommand\U{\mathcal{U}}
\newcommand\SO{\mathcal{SO}}

\newcommand\K{\mathcal{K}}
\newcommand\J{\mathcal{J}}

\newcommand\W{\mathcal{W}}
\newcommand\vphi{\varphi}

\newcommand\sS{\mathsf{S}}

\newcommand\sD{\mathsf{D}}
\newcommand\sA{\mathsf{A}}

\renewcommand\epsilon{\varepsilon}

\newcommand\rarrow{\rightarrow}

\newcommand\LieG{\mathfrak{g}}
\newcommand\LieH{\mathfrak{h}}

\newcommand\su{\mathfrak{su}}
\newcommand\so{\mathfrak{so}}
\renewcommand\sl{\mathfrak{sl}}

\renewcommand\t{\tilde}

\renewcommand\b{\bar }
\newcommand\w{\wedge}
\renewcommand\d{\partial}
\newcommand\s{\sigma}
\newcommand\bs{\boldsymbol}

\renewcommand\-{^{-1}}
\newcommand\Ad{\text{Ad}}

\renewcommand\1{\mathds{1}}

\DeclareMathOperator{\Diff}{Diff}
\DeclareMathOperator{\Aut}{Aut}
\DeclareMathOperator{\Tr}{Tr}
\DeclareMathOperator{\vol}{vol}

\newtheorem{thm}{Theorem}

\newtheorem{prop}[thm]{Proposition}
\theoremstyle{definition}

\begin{document}

\title{Dilaton from Tractor and Matter Field from Twistor}
\author{J. François${\,}^{a}$}
\date{}

\maketitle
\begin{center}
\vskip -0.5cm
\noindent
${}^a$ Service de Physique de l'Univers, Champs et Gravitation, Universit\'e de Mons -- UMONS\\
20 Place du Parc, B-7000 Mons, Belgique
\end{center}

\begin{abstract}
Tractors and twistors can be obtained via gauge reduction of the conformal Cartan geometry, thanks to the dressing field method. Perhaps surprisingly, it is possible to reduce the gauge symmetry further still. One can indeed erase Weyl symmetry thanks to the tractor field, in a way reminiscent of how the $\CC^2$-scalar field allows to erase $\SU(2)$ in the electroweak model. This  suggests an alternative to the Weyl or conformal spontaneous symmetry breaking (SSB) that some authors proposed as improvement for Standard Model physics or inflationary cosmology. Here, after gauge reduction via dressing, there remains only the Lorentz gauge symmetry and the twistor field becomes -  for all practical purposes - a Dirac spinor field. In a very simple and natural toy model, the latter acquires a mass through Lorentz SSB due to the VEV of the Weyl-invariant tractor field. 
\end{abstract}


\vspace{1.5mm}



\section{Introduction}  

Conformal and Weyl (or local scale) symmetries do not only elicit interest in exotic and/or low dimensional contexts such as string theory - where the world-sheet  of strings is required to have exact Weyl symmetry - or in the literature related to AdS/CFT holography - where a gauge theory in a bulk is dual to a conformal field theory on the boundary. They are also put forward as relevant for plain 4-dimensional physics, though usually thought of as spontaneously broken by a a so-called \emph{dilaton} field. 
For example, the spontaneous breaking of the conformal $S\!O(2, 4)$ symmetry down to a $S\!O(1,4)$ deSitter symmetry has been advocated as an alternative to cosmological inflation and as an explication of the near scale invariance of the CMB perturbations spectrum \cite{Hinterbichler&Khoury2012}. Weyl SSB as been proposed as a way to improve (or drive) inflationary cosmology, to stabilize the Standard Model (SM) and to give mass to all its fields \cite{Bars-et-al2014}. It has also been suggested that it may be the key to an improved SM where all parameters are computable, and to the solution of both the black hole information  and the firewall problems \cite{tHooft2017}. 

Model building in 4D can take advantage of conformal differential geometric frameworks such as Penrose's Twistor theory\footnote{Which was intended originally as a framework for all of physics, and short of that objective it is nevertheless  of great conceptual and technical relevance in advanced topics such as string theory \cite{Atiyah_et_al2017} and, to a lesser extent, loop quantum gravity \cite{Speziale2014}.}, or the so-called Tractor calculus  \cite{Bailey-et-al94, Gover-Shaukat-Waldron09}. Both deal only with Weyl gauge symmetry. The most general differential geometric framework to take on the full conformal symmetry is the conformal Cartan geometry \cite{Sharpe}, to which both twistor and tractor calculi are closely related \cite{Cap-Slovak09}. 
We intend to show how, within this differential geometric setup, one can  get ride of the Weyl gauge symmetry without SSB, but rather in a non-dynamical way via the \emph{dressing field method} (DFM), which is a systematic way to reduce gauge symmetries. We notice the close relation with Stora's idea about the Wess-Zumino functional  as expounded in \cite{Attard-Lazz2016} (see in particular appendix A).

The paper is organized as follows. Section \ref{Reduction of gauge symmetries via dressing} is a nutshell presentation of the dressing field method, where the main theorems regarding the gauge symmetry reduction and the residual gauge symmetries are stated in the form of propositions. Section \ref{Tractors and Twistors from gauge reduction of the conformal Cartan geometry}  describes synthetically how the tractor and twistor bundles are obtained from gauge reduction of the conformal Cartan reduction via dressing. Given these prerequisites, Section \ref{A dressing field for the Weyl  gauge symmetry} shows how a dilaton  is extracted from the tractor field and used to built a  dressing field  that allows to erase the Weyl gauge symmetry, in particular one will obtain a Weyl-invariant twistor that is identified with a Dirac spinor. This geometric construction is put to work through a very natural toy model in which the tractor fields, embedded in a potential, plays the same role as the scalar field in the electroweak model, and in which the invariant twistor is endowed with a mass through spontaneous breaking of the residual Lorentz symmetry. 
Of course this model is put forward  less as a serious proposal than 
 to stimulate the  reader's imagination and encourage him to ponder the mechanism presented. 
 Finally, in section \ref{Discussion} we gather some comments, and in particular argue that (independently of our toy model) our geometric analysis support a claim made by Jackiw and Pi a few years ago \cite{Jackiw-Pi2015} according to which  the Weyl symmetry of - at the time - recently proposed cosmological models is what they dubbed `fake'. We indicate that their comment can be replaced in a broader context within philosophy of physics regarding the distinction between artificial and substantial gauge symmetries.


\section{Reduction of gauge symmetries via dressing}
\label{Reduction of gauge symmetries via dressing}

The dressing field method  is a technically rather simple and nonetheless conceptually powerfull tool allowing to deal with gauge symmetries so as to reveal the physical degrees of freedom (d.o.f) of a theory in a way that differs markedly from both gauge fixing or spontaneous symmetry breaking  mechanisms. One can consult \cite{Attard_et_al2017} p.377 for a review with some interesting applications, and \cite{Francois2018} for an assessment of its implications regarding philosophy of physics. This section is a nutshell presentation of the method and is the occasion of fixing some notation. We refer to \cite{Attard-Francois2016_I} for the proofs of all the following assertions. 

\subsection{Invariant composite fields} 
\label{Invariant composite fields} 

The main purpose of the DFM is to propuce gauge-invariant composite fields that can adequately represents the physical d.o.f of a gauge theory. It is better formulated in the language of fiber bundle differential geometry. So let us recall what is the geometric setup of a gauge theory. 

The main object if a principal fiber bundle $\P \xrightarrow{\pi}\M$ over an $m$-dimensional spacetime $\M$ with structure Lie group $H$, also noted $\P(\M, H)$. The right action of $H$ on $\P$ is $R_hp=ph, \forall p\in \P$. 
The bundle is endowed  with a  $\LieH$-valued (Ehresmann) connection $1$-form $\omega \in \Lambda^1(\P, \LieH)$, with curvature $2$-form $\Omega \in \Lambda^2(\P, \LieH)$ given by  Cartan's structure equation $\Omega=d\omega + \tfrac{1}{2}[\omega, \omega]$. For any choice of trivialising section $\s: \U\subset \M \rarrow \P$ over an open subset of $\M$, the local representatives $A=\s^*\omega \in \Lambda^1(\U, \LieH)$ and $F=\s^*\Omega \in \Lambda^2(\U, \LieH)$ represent respectively the Yang-Mills gauge potential and its associated field-strength.
 
 Given any representations $(\rho, V)$ for $H$ one built the associated bundle $E=\P \times_\rho V$ over $\M$ whose sections $\Gamma(E)$ are in one to one correspondence with $V$-valued $\rho$-equivariant maps  $\vphi:\P \rarrow V$, i.e satisfying $R^*_h \vphi=\rho(h)^{-1}\vphi$. These represent matter fields. The connection form induces a covariant derivative on this space of maps/sections, $D\vphi=d\vphi+\rho_*(\omega)\vphi$, which represents the minimal coupling between the gauge potential and matter fields.  
 
 The diffeomorphisms of $\P$ that preserve its fibration form the group of vertical automorphisms $\Aut_v(\P):=\{\Phi \in \Diff(\P) \ | \  \pi \circ \Phi =\pi \ \&\  R_h \circ \Phi=\Phi \circ R_h  \}$. It is isomorphic to the gauge group $\H:=\{ \gamma:\P \rarrow H \ | \  R^*_h\gamma= h\- \gamma h  \}$, the isomorphism being $\Phi(p)=p\gamma(p)$. The  gauge transformations of the basic field variables are then defined as,
 \begin{align}
 \label{GT}
 \omega^\gamma&:=\Phi^*\omega=\gamma\-\omega \gamma+\gamma\-d\gamma, \qquad \Omega^\gamma:=\Phi^*\Omega=\gamma\-\Omega\gamma,  \notag\\
 \vphi^\gamma&:=\Phi^*\vphi=\rho(\gamma\-)\vphi, \quad \text{and} \quad (D\vphi^\gamma):=\Phi^*D\vphi=\rho(\gamma\-)D\vphi=D^\gamma\omega^\gamma.
 \end{align}
 
 A physical gauge theory is given by an action functional $\S=\int_\M L$, with$L=L(A, \vphi)$ a Lagrangian m-(pseudo) form on $\M$ which is required to be gauge-invariant - $L(A^\gamma, \vphi^\gamma)=L(A, \vphi)$ - or pseudo-invariant, i.e invariant up to an exact form - $L(A^\gamma, \vphi^\gamma)=L(A, \vphi)+d\alpha (A,\vphi, \gamma)$. 

 Given this geometric and gauge theoretic setup, the essential proposition of the DFM is the following.
 \begin{prop} 
 \label{Prop1}
  Let $K$ and $G$ be subgroups of $H$ such that (s.t) $K\subseteq G \subseteq H$. Note $\K\subseteq \H$ the gauge subgroup associated to $K$. Suppose there exists a map $u:\P \rarrow G$ defined by its $K$-equivariance: $R^*_ku=k\- u, \forall k\in K$, or equivalently by its $\K$-gauge transformation: $u^\gamma=\gamma\-u, \forall \gamma \in \K$. Such a map, that we call a \textbf{dressing field}, allows to construct the composite fields
 \begin{align}
 \label{CompFields}
 \omega^u&:=u\-\omega u+u\-du, \qquad \Omega^u=u\-\Omega u=d\omega^u+\tfrac{1}{2}[\omega^u, \omega^u], \notag \\
 \vphi^u&:=\rho(u\-)\vphi \quad \text{ and } \quad D^u\vphi^u=\rho(u\-)D\vphi=d\vphi^u+\rho_*(\omega^u)\vphi^u,  
 \end{align}
 which are $\K$-invariant, $K$-horizontal and therefore project (live on) the quotient subbundle $\P/K\subset \P$. 
 \end{prop}
 
 The first thing to make clear is that despite the formal resemblance, \eqref{CompFields} must not be confused with \eqref{GT}. To perform a dressing operation is not to perform a gauge transformation, a composite field is not a gauge transform of a gauge variable. This is obvious if one is attentive to the fact that a dressing field, given its defining equivariance property, cannot be an element of the gauge group ($\H$) or one of its subgroup ($\K$). As a matter of fact, the composite fields \eqref{CompFields} do not even belong to the original space of gauge variables. For example, $\omega^u$  has trivial $K$-equivariance and is $K$-horizontal, which implies that it is no longer a connection on $\P$. In particular this means that to perform a dressing is certainly not equivalent to fixing a gauge, i.e to select a single representative in the gauge orbit of any one gauge variable. 
 
 Locally, on $\U\subset \M$, keeping $u$ to denote the local dressing field (instead of $\s^*u$) and considering $\vphi$ as a section (not a map on $\P$), the local $\K$-invariant composite fields can be defined likewise: $A^u=u\-Au+u\-du$ and $\vphi^u=\rho(u\-)\vphi$. So if a $\H$-gauge theory is a priori given by $L(A, \vphi)$, but a (local\footnote{Here `local' is meant in the sense of field theory, not simply to express the fact that it is defined on $\U\in\M$. We refer to \cite{Francois2018} regarding the importance of this precision.}) $\K$-dressing field exists, then - using the invariance of the Lagrangian which hold as a formal property - one can rewrite the theory in terms of $\K$-invariant variables as  $L(A, \vphi)= L(A^u, \vphi^u)$. This shows that our theory is actually a $\H/\K$-gauge theory. Part of the gauge symmetry was spurious and could be eliminated by a change of field variables.

\subsection{Residual gauge symmetries} 
\label{Residual gauge symmetries} 

The next interesting question concerns the residual gauge symmetries exhibited by the composite fields. For this question to even make immediate sense it is necessary to assume that $K$ is not merely a subgroup but a normal subgroup of $H$, $K\triangleleft H$, so that the quotient $H/K$ is a group in its own right, that we note $J$ for convenience. The quotient subbundle $\P/K$ is then a $J$-principal bundle $\P'(\M, J)$ on which live the composite fields \eqref{CompFields}.  Note $\J$ the associated gauge group. 

The action of $\J$ on the original gauge variables is known, so its action on the composite fields obviously depends on how it affects the dressing field. The first nice possibility is as follows. 

\begin{prop}
\label{Prop2}
Suppose the $\K$-dressing field $u$ has $J$-equivariance given by $R^*_ju=\Ad_{j\-} u$, $\forall j\in J$, or equivalently has $\J$-gauge transformation $u^{\gamma'}={\gamma'}\- u\gamma', \forall \gamma'\in \J$. Then $\omega^u$ is a $J$-principal connection on $\P'$ with curvature $\Omega^u$, and $\vphi^u$ is a $J$-equivariant map on $\P'$ acted upon by the covariant derivative $D^u=d +\rho_*(\omega^u)$. The $\J$-gauge transformation of the $\K$-invariant composite fields is then as in \eqref{GT},
\begin{align}
\label{ReSym1}
(\omega^u)^{\gamma'} &={\gamma'}\- \omega^u \gamma' + {\gamma'}\-d\gamma',
\qquad (\Omega^u)^{\gamma'}={\gamma'}\- \Omega^u \gamma', \notag \\
(\vphi^u)^{\gamma'}&= \rho\left({\gamma'}\-\right)\vphi^u \quad \text{ and } 
\quad (D^u\vphi^u)^{\gamma'}=\rho\left({\gamma'}\-\right)D^u\vphi^u.
\end{align}
\end{prop}
\noindent The composite fields are standard gauge fields, albeit for a smaller gauge symmetry group, and are amenable to all manner of usual theoretic tortures. 
\bigskip

The next intriguing possibility is a variation on this. Consider a map $C:\P'\times J \rarrow G$, $(p, j) \mapsto C_p(j)$, satisfying $C_p(jj')=C_p(j)C_{pj}(j')$. It induces a map $C(\gamma'): \P' \rarrow G$, $p \mapsto C_p\left(\gamma'(p)\right)$.
\begin{prop}
\label{Prop3}
Suppose the $\K$-dressing field $u$ has $J$-equivariance given by $R^*_j u=j\- u C(j), \forall j\in J$, or equivalently has $\J$-gauge transformation $u^{\gamma'}= {\gamma'}\- u C(\gamma'), \forall \gamma'\in \J$.  Then $\omega^u$ is a generalized connection (call it a twisted connection) whose curvature $\Omega^u$ is $\Ad_{C(J)}$-tensorial, and $\vphi^u$ is a $\rho\left(C(J)\right)$-equivariant map on $\P'$ acted upon by the covariant derivative $D^u=d +\rho_*(\omega^u)$. The $\J$-gauge transformation of the composite fields are then, 
\begin{align}
\label{ReSym2}
(\omega^u)^{\gamma'} &=C(\gamma')\- \omega^u C(\gamma') + C(\gamma')\-dC(\gamma'),
\qquad (\Omega^u)^{\gamma'}=C(\gamma')\- \Omega C(\gamma'), \notag \\
(\vphi^u)^{\gamma'}&= \rho\left(C(\gamma')\-\right)\vphi^u \quad \text{ and } 
\quad (D^u\vphi^u)^{\gamma'}=\rho\left(C(\gamma')\-\right)D^u\vphi^u.
\end{align}
\end{prop}

\noindent The composite fields behaves as gauge fields of a new kind, implementing the gauge principle of field theory. 
\medskip

In both situations described, the philosophy of the DFM  still  applies, meaning we might want and be able to further reduce the residual gauge symmetry at hand. 

\begin{prop}
\label{Prop4}
Suppose indeed that a $\J$-dressing field is available: $u':\P' \rarrow J$, satisfying $R^*_ju'=j\- u', \forall j\in J$, or ${u'}^{\gamma'} ={\gamma'}\-u', \forall \gamma'\in \J$. It can be used to dress the composite fields \eqref{CompFields} and compensate for the transformation \eqref{ReSym1}. In order not to spoil the $\K$-invariance obtained after the first dressing, $u'$ must satisfy the compatibility condition $R^*_k u'=u', \forall k\in K$, or equivalently ${u'}^\gamma=u', \forall \gamma \in \K$. 

Given such $u'$, the map $C(u'):\P' \rarrow G$ is a $C(J)$-dressing field satisfying $R^*_jC(u')=C(j)\-C(u')$ or equivalently $C(u')^{\gamma'}=C(\gamma')\-C(u')$. It can be used to dress the composite fields \eqref{CompFields} and compensate for the transformation \eqref{ReSym2}. In order not to spoil the $\K$-invariance obtained after the first dressing, the map $C(j):\P' \rarrow G$ must satisfy $R^*_kC(j)=C(j)$, or equivalently $C(j)^\gamma=C(j)$.  

In both situations, the composite fields $\chi^{uu'}=(\chi^u)^{u'}$ or $\chi^{uC(u')}=(\chi^u)^{C(u')}$, with $\chi=\{\omega, \Omega, \vphi \}$, are  $\H$-invariant.
\end{prop}


\section{Tractors and Twistors from gauge reduction of the conformal Cartan geometry} 
\label{Tractors and Twistors from gauge reduction of the conformal Cartan geometry} 

It can be  shown that the Tractor and Twistor bundles can be obtained as end-products of gauge reduction via dressing of associated bundles of the conformal Cartan geometry, and that tractor and twistor connections are really just representations of the dressed (normal) Cartan connection. As a matter of fact the dressing field approach has two advantages: First, it allows to generalize seamlessly tractor and twistor calculi to conformal manifolds with torsion. Secondly, it reveals that tractors and  twistors are not sections of usual associated bundles coming from a representation of the structure group of a principal bundle, but that they are representatives of gauge fields of a non-standard kind as they are sections of twisted bundles constructed not through a representation but through a twisting map $C$ as described above. 
In the following we describe these results as synthetically as possible,  and refer back to \cite{Attard-Francois2016_I, Attard-Francois2016_II} 
 for further details. 

We work over a $4$-dim  (spin) manifold $\M$. The conformal Cartan geometry is the bundle $\P(\M, H)$ whose structure group  is a subgroup of the conformal group $S\!O(2,4)$ - which preserves the metric $\Sigma=\begin{psmallmatrix} 0&0&-1 \\ 0&\eta&0 \\-1 &0&0 \end{psmallmatrix}$, where $\eta$ is the Minkowski metric with signature $(+---)$. It is given by  \cite{Sharpe}
\begin{align*}
 H = K_0\ltimes K_1=\left\{ \begin{psmallmatrix} z &  0 & 0  \\  0  & S & 0 \\ 0 & 0 & z^{-1}  \end{psmallmatrix}  \begin{psmallmatrix} 1 & r & \sfrac{1}{2}rr^t \\ 0 & \1_4 & r^t \\  0 & 0 & 1\end{psmallmatrix}  \bigg|\ z\in\RR_*^+,\ S\in S\!O(1, 3), 
\ r\in \RR^{4*} \right\}.
\end{align*} 
One has $r^t = (r \eta^{-1})^T$ (the operation ${}^T\,$ is the usual matrix transposition). The group $K_1$ is the abelian group of conformal boost, $\RR^{4*}$ is the dual of $\RR^4$. This bundle is endowed with a Cartan connection, a $\LieG$-valued $1$-form, where $\LieG=\LieG_{-1}+\LieG_0+\LieG_1=\so(2,4)$, $[\LieG_i, \LieG_j] \subset \LieG_{i+j}$, is the graded Lie algebra of the conformal group.
  In matrix representation,
\begin{align*}
\mathfrak{g} = \left\{ 
\begin{psmallmatrix} \epsilon &  \iota & 0  \\  \tau  & s & \iota^t \\ 0 & \tau^t & -\epsilon  \end{psmallmatrix} \bigg|\ \left\{\epsilon \in \RR, s \in \so(1, 3)\right\} \simeq \LieG_0,\ \ \iota\in\mathbb{R}^{4*}\simeq \LieG_1, \  \tau\in\mathbb{R}^4 \simeq \LieG_{-1}  
\right\} 
\supset
\LieH = \left\{ \begin{psmallmatrix} \epsilon &  \iota  & 0  \\  0  & s & \iota^t \\ 0 & 0 & -\epsilon  \end{psmallmatrix} \right\}.
\end{align*} 
 Working locally,  the Cartan connection $\varpi \in \Lambda^1(\U, \LieG)$ and its curvature $\Omega$ have the matrix form $\varpi =\begin{psmallmatrix} a & P & 0 \\ \theta & A & P^t \\0 & \theta^t & -a \end{psmallmatrix}$ and $\Omega=d\varpi+\varpi^2=\begin{psmallmatrix} f & C & 0 \\ \Theta & W & C^t \\[0.2mm]0 & \Theta^t & -f \end{psmallmatrix}$.
The canonical representation of $H$ allows to built the real vector bundle $E=\P\times_H \RR^6$ whose sections are (equivalent to) $H$-equivariant maps on $\P$ with local expression
$\vphi: \U \subset \M \rarrow \RR^6$, given explicitly as column vectors $\vphi=\begin{psmallmatrix} \rho\\ \ell \\ \s  \end{psmallmatrix}$,  where $ \ell=\ell^a \in \RR^4$, $\rho \in \RR$ and $\s \in \RR_*$.
The covariant derivative on such maps is $D\vphi=d\vphi+\varpi\vphi$. 

 There is a complex representation of $H$, from the double cover group morphism $\b H \xrightarrow{2:1} H$, with
 \begin{align}
 \b H=\b K_0 \ltimes \b K_1 := \left\{ \begin{psmallmatrix} z^{\sfrac{1}{2}} {\b S}^{-1*} & 0 \\ 0 & z^{-\sfrac{1}{2}} \b S\  \end{psmallmatrix}\begin{psmallmatrix} \1_2 & -i \b r \\ 0 & \1_2 \end{psmallmatrix} \bigg|\ z\in\RR_*^+,\ \b S\in S\!L(2, \CC), 
\ \b r\in \text{Herm}(2, \CC)   \right\}.
\end{align}
  It is induced  by the vector space isomorphism $\RR^4 \rarrow$ Herm$(2, \CC)=\{M \in M_2(\CC) \ | \ M^*=M \}$, $x \mapsto \b x=x^a\s_a$ - where ${}^*$ means transconjugation, $\s_0=\1_2$ and $\s_{k=\{1,2,3\}}$ are Pauli matrices - and  by the double covering group morphism Spin$(1,3)\simeq S\!L(2, \CC) \xrightarrow{2:1} S\!O(1,3)$, $\b S \mapsto S$. This also induces the double cover  Spin$(2,4)\simeq S\!U(2,2) \xrightarrow{2:1} S\!O(2,4)$, where $S\!U(2, 2)$ is the group preserving the metric $\b\Sigma=\begin{psmallmatrix} 0 & \1_2 \\ \1_2 & 0  \end{psmallmatrix}$, which in turn implies the Lie algebra isomorphism $\so(2,4)=\LieG \rarrow \su(2,2)=\b \LieG$ with
\begin{align*}
\label{LieAlg-Iso}
\b\LieG = \left\{ 
\begin{psmallmatrix} -(\b s^* -\sfrac{\epsilon}{2}\1) &  -i\,\b\iota \\  i\b\tau  & \b s-\sfrac{\epsilon}{2}\1  \end{psmallmatrix}   
\bigg|\ \epsilon\in\RR, \b s\in \sl(2,\CC) \text{ and }  \b\iota, \b \tau \in \text{Herm}(2, \CC)  \right\}.
\end{align*}
The Cartan connection  then has the complex representation 
$\b\varpi=\begin{psmallmatrix}  -( \b A^* - \sfrac{a}{2}\1 )  &  -i\b P \\ i\b \theta & \b A -\sfrac{a}{2}\1  \end{psmallmatrix}$,  and similarly  its curvature is   $\b\Omega=\begin{psmallmatrix}  -( \b W^* - \sfrac{f}{2}\1 )  &  -i\b C \\ i\b \Theta & \b W -\sfrac{f}{2}\1  \end{psmallmatrix}$.
 The associated complex vector bundle $\sf{E}=\P\times_{\b H} \CC^4$ has sections equivalent to  $\ H$-equivariant maps on $\P$ with local expression $\psi:\U \subset \M \rarrow \CC^4$, given explicitly as $\psi=\begin{psmallmatrix} \pi\\[1mm] \omega \end{psmallmatrix}$, where $\pi$ and $\omega \in\CC^2$ are dual Weyl spinors. The covariant derivative on such maps is $\b D\psi=d\psi +\b \varpi \psi$.
\bigskip

Despite what a naive approach would have led us to think, $E$ is not the tractor bundle and $\sf{E}$ is not the twistor bundle. This is clear when one inspects the gauge transformations of $\vphi$ and $\psi$ under $\K_0=\SO \times \W \subset \H$ - where $\W=\left\{Z:\U \rarrow W\subset K_0 \ |\   Z=\begin{psmallmatrix} z & 0 & 0 \\ 0 & \1_4 & 0 \\ 0  & 0 &z\- \end{psmallmatrix} \right\}$ is the group of Weyl rescalings - and compare with the transformation laws of tractors and twistors in the literature: they are nothing alike. Curiously,  the correct laws are recovered only after a gauge symmetry reduction via dressing.
\medskip

Indeed, from the Cartan connection itself one builds a (local!) $\K_1$-dressing field $u_1 :\U\rarrow K_1$, explicitly  $u_1=\begin{psmallmatrix}  1 & q & \sfrac{1}{2}qq^t \\ 0 & \1 & q^t \\ 0 & 0 & 1 \end{psmallmatrix}$ with $q=q_a=a_\mu{e^\mu}_a$, where  $a=a_\mu dx^\mu$ and  $\theta^a={e^a}_\mu dx^\mu$ is the soldering form whose components are the tetrad field. From the $\K_1$-gauge transformation of $\varpi$ one checks that $u_1^{\gamma_1}=\gamma_1\-u_1, \forall \gamma_1\in \K_1$, as is defining of a dressing field. One then simply apply Proposition \ref{Prop1} and built the $\K_1$-invariant composite fields
\begin{equation}
\begin{split}
\label{K1CompFields}
\varpi_1 &:=u_1\- \varpi u_1 +{u_1}\-du_1, \qquad \Omega_1=d\omega_1 + {\omega_1}^2,   \\  
\vphi_1 &:=u_1\-\vphi, \quad \text{and} \quad D_1\vphi_1=d\vphi_1 + \varpi_1 \vphi_1,  \\
\psi_1&:={\b u_1}^{-1} \psi, \quad \text{and} \quad \b D_1 \psi_1=d\psi_1 +\b\varpi_1 \psi_1. 
\end{split}
\end{equation}
Their residual gauge symmetry group is obviously $\K_0$, but what is the explicit transformation law? The undressed field variables being standard gauge fields, we only need to find the $\K_0$ transformation of the dressing  field $u_1$.

It turns out that given the gauge transformation of $\varpi$ under $\SO=\left\{\sf{S}:\U \rarrow \sf{SO}(1,3)\ | \ \sf{S}=\begin{psmallmatrix} 1 & 0 & 0 \\ 0 & S & 0 \\ 0 & 0 & 1 \end{psmallmatrix}  \right\} $, one finds that $u_1^{\sf{S}}=\sf{S}\- u_1 \sf{S}$. This is an instance of Proposition \ref{Prop2}, so one concludes that the above composite fields 
are genuine standard $\SO$-gauge  fields with transformations given  by the usual relations \eqref{ReSym1}.

Following  the same logic by considering next the gauge transformation of $\varpi$ under $\W$, one easily finds that $u_1^Z=Z\- u_1 C(z)$ where the twisting map $C(z):\U \subset \M \rarrow K_1W\subset H$ is explictly given as $C(z):=
\begin{psmallmatrix} z \ & \Upsilon(z)\ &  \sfrac{z\-}{2} \Upsilon(z) ^2 \\ 0 & \1_4 & z\-\Upsilon(z)^t \\ 0 & 0 & z\-  \end{psmallmatrix}$, with $\Upsilon(z)=\Upsilon(z)_a:=z\-\d_\mu z \ {e^\mu}_a$, and 
$\Upsilon(z)^2=
\Upsilon(z) \Upsilon(z)^t$.
Notice that $C(z')C(z)\neq C(z'z)$, but instead one has $C(z'z)=C(z')\, {Z'}\-C(z)Z'$. It is a particular case of Proposition \ref{Prop3}, so one concludes that the composite fields \eqref{K1CompFields} are non-standard $\W$-gauge fields with transformations given by \eqref{ReSym2}. Furthermore,  the relation $\vphi_1^Z=C(z)\- \vphi_1$ is precisely the right rescaling law for a tractor field, and $\psi_1^Z= {\b C(z)}\-\psi_1$ - with $\b C(z)=\begin{psmallmatrix} z^{\sfrac{1}{2}} \1_2\ \ &  -i\ z^{-\sfrac{1}{2}}\b \Upsilon(z) \\ 0 & z^{-\sfrac{1}{2}}\1_2 \end{psmallmatrix}$ - is the rescaling law for a  twistor. 

The immediate implication is that $\varpi_1$ is a generalization of the so-called \emph{tractor connection}, and that $\b\varpi_1$ is likewise a generalization of the \emph{twistor connection}, since their torsion $\Theta$ is non zero. Only when one dresses the (unique) \emph{normal} Cartan connection $\varpi_N$ - depending only on the d.o.f of $\theta$ 
 \cite{Sharpe} - do we end-up with the usual tractor $\varpi_{N,1}$ and twistor $\b\varpi_{N,1}$ connections. Explicitly, 
$\varpi_{N,1}=\begin{psmallmatrix} 0 & P_1 & 0  \\ \theta  & A_1  & P^t  \\ 0 & \theta^t & 0  \end{psmallmatrix}$ with $\theta$ the soldering form, $A_1$ the Lorentz/spin connection, and $P_1$  the Schouten $1$-form. The curvature is 
$\Omega_{N, 1}=\begin{psmallmatrix} 0& C_1& 0\\ 0 & W_1 & C_1^t \\ 0 & 0 & 0  \end{psmallmatrix}$, with $W_1$ the Weyl $2$-form and $C_1$ the Cotton $2$-form. Again, we refer to \cite{Attard-Francois2016_I, Attard-Francois2016_II}
  for a more detailed account.

Notice that the dressed Cartan connection induces a conformal class of metric $c=[g]$ on $\M$ via its soldering part. Indeed if $\varpi_1$ induces, via $\theta$, the metric $g:=e^T\eta e$, then $\varpi_1^Z$ induces via $\theta^Z=z\theta$ the metric $z^2g$. Nevertheless it is the normal Cartan geometry $(\P, \varpi_{N,1})$ which is truly equivalent to the data of the conformal manifold $(\M, c)$.

\section{A dressing field for the Weyl  gauge symmetry}
\label{A dressing field for the Weyl  gauge symmetry}

So far so good. Interestingly, it is possible to perform still another gauge reduction on this tractor-twistor geometric set-up: This time we can erase the Weyl gauge symmetry.

The explicit Weyl rescaling law for the tractor field $\vphi_1=\begin{psmallmatrix} \rho_1 \\ \ell_1 \\ \s \end{psmallmatrix}\ $ is:  $\vphi_1^Z =C(z)\- \vphi_1 = \begin{psmallmatrix} z\-\left( \rho_1 - \Upsilon(z)\ell_1 + \sfrac{\s}{2} \Upsilon(z)^2 \right) \\ \ell_1 - \Upsilon(z)^t \s \\[1mm] z\s \end{psmallmatrix}$. Remark that the trator field supplies a candidate dressing field for the Weyl gauge symmetry, otherwise known as a dilaton. Indeed defining $\phi:=\s\-$, we have $\phi^z=z\-\phi, \forall\  z:\!\U\rarrow\RR_*^+$, which is the defining property of a dressing field. It is by the way clear   that $\phi^{\gamma_1}=\phi$. Also, $C(z)$ depends only on $e={e^a}_\mu$ which satisfies $e^{\gamma_1}=e$, so that $C(z)^{\gamma_1}=C(z)$. Therefore, by Proposition \ref{Prop3} the map $C(\phi):\U \rarrow K_1W$, given by $C(\phi)=\begin{psmallmatrix} \phi \ & \Upsilon(\phi)\ &  \sfrac{\phi\-}{2} \Upsilon(\phi) ^2 \\ 0 & \1_4 & \phi\-\Upsilon(\phi)^t \\ 0 & 0 & \phi\-  \end{psmallmatrix}$ with  $\Upsilon(\phi)=\Upsilon(\phi)_a:=\phi\-\d_\mu \phi \ {e^\mu}_a$, satisfies both $C(\phi)^\W=C(z)\-C(\phi)$ and $C(\phi)^{\gamma_1}=C(\phi)$. It is thus a dressing field adequate for the purpose of erasing the residual Weyl gauge symmetry of the composite fields  \eqref{K1CompFields} while preserving their conformal boost invariance. Applying then again  Proposition \ref{Prop1} and \eqref{CompFields}, one defines the following $\K_1$- \emph{and} $\W$-invariant composite fields
\begin{equation}
\label{WCompFields}
\begin{split}
\bs\varpi&:=C(\phi)\- \varpi_1 C(\phi) + C(\phi)\-dC(\phi), \quad \bs \Omega =d\bs\varpi + \bs\varpi^2,\\
\bs \vphi&:=C(\phi)\- \vphi_1, \qquad \text{and} \qquad  \bs{D\vphi}=d\bs\vphi + \bs{\varpi\vphi}, \\
\bs \psi&:= \b C(\phi)\- \psi_1, \qquad \text{and} \qquad  \b{\bs D}\bs\psi=d\bs\psi + \b{\bs\varpi}\bs\psi.
\end{split}
\end{equation}
We have explicitly 
$\bs\varpi=\begin{psmallmatrix} 0  & \bs P & 0 \\[0.1mm]   \bs\theta & \bs A & \bs{P}^t\\[0.2mm]  0 & \bs{\theta}^t & 0 \end{psmallmatrix}
= \begin{psmallmatrix} 0\ & \phi\- \left(P_1+\nabla\Upsilon(\phi) - \Upsilon(\phi)\theta \Upsilon(\phi) +\sfrac{1}{2}\Upsilon(\phi)^2\theta^t\right) \  & 0 \\[0.5mm] \phi\theta \ \ & A_1+\theta \Upsilon(\phi) - \Upsilon(\phi)^t\theta^t & * \\[0.5mm]  0\ & \phi\theta^t & 0 \end{psmallmatrix}$,
which  induces via its Weyl-invariant soldering form $\bs\theta$ the invariant Lorentzian metric $\bs g=\phi^2g$ on $\M$.
Also, the invariant tractor field is now $\bs\vphi=\begin{psmallmatrix} \bs \rho \\ \bs\ell \\ 1 \end{psmallmatrix}$, and has manifestly one less d.o.f than its undressed counterpart.

 Quite obviously, there now remains only the Lorentz gauge group $\SO$. The question is how the fields \eqref{WCompFields} behaves w.r.t this residual symmetry. Since it is already established that the $\K_1$-invariant fields \eqref{K1CompFields} are genuine standard $\SO$-gauge fields, it is only necessary to consider the Lorentz transformation of the dressing field $C(\phi)$. It is easily checked that $C(\phi)^\sS=\sS\-C(\phi)\sS$, so by Proposition \ref{Prop2} the $\K_1$- and $\W$-invariant composite fields \eqref{WCompFields} are genuine standard $\SO$-gauge fields, therefore transforming according to \eqref{ReSym1}:
 \begin{equation}
 \begin{split}
 \bs\varpi^\sS &= \sS\- \bs\varpi \sS +\sS\-d\sS,\qquad \bs\Omega^\sS=\sS\-\bs\Omega \sS, \\
 \bs\vphi^\sS &= \sS\-\bs\vphi,\qquad \text{and} \qquad (\bs{D\varphi})^\sS=\sS\-\bs{D\vphi}, \\
 \bs\psi^\sS &= {\b\sS}\-\bs\psi, \qquad \text{and} \qquad (\b{\bs D}\bs\psi)^\sS={\b\sS}\- \b{\bs D}\bs\psi.
 \end{split}
 \end{equation}

At this point, several interesting observations can be made. First, it is striking that the Weyl-dressed twistor $\bs\psi$, which now only supports Lorentz gauge transformations via $\bar\SO=\left\{ \b\sS:\U \rarrow S\!L(2, \CC)\oplus S\!L(2,\CC)^* \ |\  \b S=\begin{psmallmatrix}  {\b S}^{-1*}& 0 \\ 0 &  \b S\end{psmallmatrix} \right\}$, can then be  interpreted as a Dirac spinor field. 

 Furthermore, let us highlight - even if that might seem obvious to some readers - that the underlying geometric set-up of the twistor sector provides  a preferred choice of basis for gamma matrices. Indeed, the Lie algebra isomorphism $\so(2,4) \rarrow \su(2,2)$ necessary for the construction of the twistor sector induces the vector space isomorphism $\RR^4 \rarrow$  
$\ \b\LieG_{-1}\oplus \b\LieG_1 \subset \su(2,2)$, given explicitly by $x=x^a b_a \mapsto \begin{psmallmatrix} 0 & i\b{x^t} \\ i \b x & 0  \end{psmallmatrix}= \tfrac{i}{\sqrt{2}} x^a \begin{psmallmatrix} 0 & \t\s_a \\ \s_a & 0 \end{psmallmatrix}=\tfrac{i}{\sqrt{2}} x^a\bs{\gamma}_a$ \label{iso}. The gamma matrices  $\bs{\gamma}_a:=\left\{ \bs{\gamma}_0=\begin{psmallmatrix} 0 & \s_0 \\  \s_0 & 0 \end{psmallmatrix}, \ \bs{\gamma}_k=\begin{psmallmatrix} 0 & -\s_k \\ \s_k & 0 \end{psmallmatrix}_{|k=1,2,3} \right\}$ are otherwise known as the Weyl - or chiral - basis. From this isomorphisms follows that they Lorentz transform via $\bs\gamma_a^\sS={(S\-)^b}_a \bs\gamma_b=\b\sS \- \bs\gamma_a \b\sS$. As it should, they satisfy the Clifford algebra relation  $\{ \bs\gamma_a,\bs\gamma_b  \}=2\eta_{ab} \1_4$. The hermiticity properties $\bs\gamma_0^*=\bs\gamma_0$ and $\bs\gamma_k^*=-\bs\gamma_k$, are here necessary and do not result from an arbitrary convenient choice (as is sometimes the case in the physics literature). The fact that $\tfrac{1}{2}[\bs\gamma_a \bs\gamma_b]$ - the so-called \emph{spin operator} $\s_{ab}$ \cite{Pollock2010}  - is a basis of $\sl(2, \CC)\oplus\sl(2,\CC)^*\!=\b\LieG_0$ simply reflects the graded structure of $\su(2,2)=\b\LieG$: $[\b\LieG_{-1}, \b\LieG_1] \subset \b\LieG_0$. 
The fifth gamma matrix  $\bs\gamma_5:=i\bs\gamma_0\bs\gamma_1\bs\gamma_2\bs\gamma_3=\begin{psmallmatrix} \1_2 & 0  \\ 0 & -\1_2 \end{psmallmatrix}$  satisfies $\bs\gamma_5^*=\bs\gamma_5$.
By the way, $\bs\gamma^a:=\eta^{ab}\bs\gamma_b=\begin{psmallmatrix} 0 & \s_a \\ \t\s_a & 0 \end{psmallmatrix}=\left\{ \bs{\gamma}^0=\bs\gamma_0=\begin{psmallmatrix} 0 & \s_0 \\  \s_0 & 0 \end{psmallmatrix}, \ \bs{\gamma}^k=-\bs\gamma_k=\begin{psmallmatrix} 0 & \s_k \\ -\s_k & 0 \end{psmallmatrix}_{|k=1,2,3} \right\}$, so $\bs\gamma^5=-\bs\gamma_5$. As is then usual, one defines the chiral projectors $P_L:=\tfrac{1}{2}(\1_4 + \bs\gamma_5)$ and $P_R:=\tfrac{1}{2}(\1_4 - \bs\gamma_5)$, so that $\bs\psi_L=P_L\bs\psi=\begin{psmallmatrix} \bs\pi \\ 0 \end{psmallmatrix}$ and  $\bs\psi_R=P_R\bs\psi=\begin{psmallmatrix} 0 \\ \bs\omega \end{psmallmatrix}$. By a slight abuse of notation one can then rewrites the Weyl-invariant twistor like a Dirac spinor $\bs\psi=\begin{psmallmatrix} \bs\psi_L \\ \bs\psi_R \end{psmallmatrix}$.

Notice that the
 Lorentz part of the dressed Cartan connection $\bs\varpi$, $\bs \sA=\begin{psmallmatrix} 0 & 0&0\\0&\bs A& 0\\0&0&0 \end{psmallmatrix}$, maps to the complex representation $\b{\bs\sA}=\begin{psmallmatrix} -\b{\bs A}^* &0\\0& \b{\bs A} \end{psmallmatrix}$. We have therefore the Lorentz covariant derivative acting on spinors: ${\bs\sD}\bs\psi:=d\bs\psi+\b{\bs \sA}\bs\psi$, as a piece of the invariant Cartan derivative $\b{\bs D}\bs\psi$. 
Defining the gamma $1$-form $\bs\gamma=\bs\gamma_a \bs\theta^a=\bs\gamma_a {\bs{e}^a}_\mu dx^\mu=:\bs\gamma_\mu dx^\mu$ - where $\bs\gamma_\mu$ are  sometimes called ``curved-space gamma matrices'' and satisfy $\{\bs\gamma_\mu, \bs\gamma_\nu \}=2\bs{g}_{\mu\nu}\1_4$ - 
one can build the Dirac operator $m$-form $\bs{\slashed{\sD}}\bs\psi=\bs\gamma \w *_{\bs g}\bs\sD\bs\psi =\bs\gamma^\mu \bs\sD_\mu\bs\psi \vol$, where $*_{\bs g}$ is the Hodge star operator associated with the invariant Lorentzian metric $\bs g$ and $\vol=\sqrt{|\bs g |}dx^4$ is the volume form on $\M$. 
Thanks to  the  group metric $\b\Sigma$ of $S\!U(2,2)$, one can then naturally write a Lagrangian for the (massless) Dirac  field $\bs\psi$: $L_{\text{\tiny{Dirac}}}(\bs\psi, \b{\bs\sA})=\langle \bs\psi, \bs{\slashed{\sD}}\bs\psi \rangle= \bs\psi^* \b\Sigma \bs{\slashed{\sD}}\bs\psi=\b{\bs\psi} \bs{\slashed{\sD}}\bs\psi$. Notice that the usual definition for the Dirac adjoint spinor $\b{\bs\psi}=\bs\psi^*\bs\gamma^0$ - initially suggested by Bargmann instead of the problematic curved-space version $\bs\psi^*\bs\gamma^a(\bs{e}\-)_a^{\ 0}$ \cite{Bargmann1932} - is automatic here due to  the welcomed fact that $\bs\gamma^0=\b\Sigma$.%
\footnote{An incidental fact that should not obscure the conceptual distinction that $\bs\gamma_a$ are linear operators on the space of spinors while $\b\Sigma$ defines an hermitian form on it. This speaks in favor of the view proposed here.}
\medskip

Now, what can be done with all these ingredients at our disposal? Well, it is a striking fact that the  $\SU(2)\times\U(1)$ electroweak model can be treated via the DFM: A dressing field is extracted from the $\CC^2$-scalar field, allowing to erase  $\SU(2)$ and leaving only a $\U(1)$ residual gauge symmetry. The (dressed) scalar field embedded in a potential gives mass to the (dressed) gauge and matter fields via its (unique) nonvanishing VEV. The notion of SSB never appears here. See \cite{Francois2014, Attard_et_al2017} for a  technical account and \cite{Francois2018} for a philosophical discussion of this alternative approach. Given then the striking geometrical analogy with the present situation, one could suggest writing the most natural Lagrangian for the conformal Cartan geometry that would imitate the electroweak model: 
\begin{align}
\label{H}
L(\varpi, \vphi, \psi)= \tfrac{1}{2}\!\Tr( \Omega \w *_{\bs g}\Omega ) + \langle D\vphi, *_{\bs g} D\vphi \rangle - V(\vphi) + \langle \psi, *_{\bs g}\b{\slashed{D}} \psi \rangle - |\vphi| \langle \psi, *_{\bs g}\psi \rangle.
\end{align}
The $\RR^6$-field $\vphi$ is the analogue to the $\CC^2$-scalar field  and is embedded in the potential $V(\vphi)=\left(  \alpha \langle \vphi, \vphi \rangle + \beta \langle \vphi, \vphi  \rangle^2\right)\,\vol$, where for the usual reasons $\alpha$ is an unconstrained real parameter while $\beta >0$. The last term is a Yukawa-type coupling where $|\vphi|^2:=\langle \vphi, \vphi \rangle=\vphi^T \Sigma \vphi$. 

A priori this Lagrangian describes a conformal $\H= \K_0\rtimes \K_1$-gauge theory. But  according to our discussion following Proposition \ref{Prop1}, given that there is a $\K_1$-dressing field $u_1$, the Lagrangian can be rewritten as
\begin{align}
\label{K0}
L(\varpi_1, \vphi_1, \psi_1)= \tfrac{1}{2}\!\Tr( \Omega_1 \w *_{\bs g}\Omega_1 ) + \langle D_1\vphi_1, *_{\bs g} D_1\vphi_1 \rangle - V(\vphi_1) + \langle \psi_1, *_{\bs g}\b{\slashed{D}}_1 \psi_1 \rangle - |\vphi_1| \langle \psi_1, *_{\bs g}\psi_1 \rangle,
\end{align}
which is a $\K_0$-gauge theory involving the tractor field $\vphi_1$ and the twistor field $\psi_1$, both a priori massless. 

But it is not over yet. Given that we also have a dressing field for the Weyl symmetry, the `super-dilaton' $C(\phi)$ (with value in $K_1W \subset H$), the Lagrangian is further rewritten as
\begin{align}
\label{Lorentz}
L(\bs\varpi, \bs\vphi, \bs\psi)= \tfrac{1}{2}\!\Tr( \bs\Omega \w *_{\bs g}\bs\Omega ) + \langle \bs{D}\bs\vphi, *_{\bs g} \bs{D\vphi} \rangle - V(\bs\vphi) + \langle \bs\psi, *_{\bs g}\b{\slashed{\bs D}} \bs\psi \rangle - |\bs\vphi| \langle \bs\psi, *_{\bs g}\bs\psi \rangle.
\end{align}
It is a $\SO$-gauge theory involving a Weyl-invariant tractor $\bs\vphi$ 
 and the Dirac spinor $\bs\psi$, both coupled to a Weyl-invariant conformal-type gravitational field $\bs\varpi$ whose dynamics is given by the Yang-Mills-type first term.\footnote{The latter reduces to the standard Weyl squared gravity $\tfrac{1}{2}\!\Tr( \bs W \w *_{\bs g}\bs W)$ when we restrict to the normal Cartan geometry. }
The potential function $V:\RR^6 \rarrow \RR$ has differential (in the sense of multivariable calculus) $dV_{\bs\vphi}=2\left(\alpha + 2 \beta\langle\bs\vphi, \bs\vphi\rangle\right) \langle \bs\vphi|$. The potential is thus extremal either for $\bs\vphi=0$ or for $\bs\vphi_0$ s.t  $\langle\bs\vphi_0, \bs\vphi_0\rangle=-\tfrac{\alpha}{2\beta}$. The minimum of the potential is reached for the latter, which is then the so-called vacuum expectation value (VEV).
The invariant tractor $\bs\vphi$ therefore endows both the gravitational and matter fields with masses through its non vanishing VEV. In particular the Dirac spinor $\bs\psi$ has a mass $m=\sqrt{-\tfrac{\alpha}{2\beta}}$. 

Notice that this 
 is achieved not by breaking the Weyl symmetry, which is \emph{erased} via dressing, but via spontaneous \emph{Lorentz} symmetry breaking. 
 A phenomenon for which there is also room for in some models describing the Planck era of the early Universe in string theory, loop quantum cosmology, and even the more conservative Standard Model Extension (SME) , and whose low energy relics could be within experimental reach \cite{collins_perez_sudarsky_2009,Colladay-Kostelecky1998, Mattingly2005}.
  Since Lorentz violation is related to CPT violation \cite{Greenberg2002},
  it could be part of the explanation of the matter-antimatter asymmetry in the Big Bang baryogenesis \cite{Bertolami-et-al1997, Carroll-Shu2006}.%
\footnote{In this respect, one remarks that another natural term that could be added in the Lagrangian \eqref{Lorentz} (but not in \eqref{H} and \eqref{K0} since it would explicitly break the $\K_1$ and $\W$ symmetries) is $-\langle \bs\psi, \b{\bs\vphi} \bs\psi \rangle$, where the invariant tractor $\bs\vphi$ is mapped to $\b{\bs\vphi}:=\begin{psmallmatrix} \bs\rho \1_2 &  i \b{\bs\ell^t} \\ i\b{\bs\ell} & \1_2 \end{psmallmatrix}$ so that $\sS^{-1}\bs\vphi \mapsto \b\sS^{-1} \b{\bs\vphi} \b\sS$. It contains the term $\tfrac{-i}{\sqrt{2}} \bs\ell^a \langle \bs\psi, \bs\gamma_a \bs\psi \rangle=\tfrac{-i}{\sqrt{2}} \ell^a \b{\bs\psi} \bs\gamma_a \bs\psi$ which is  the kind of Lorentz and CPT breaking terms appearing in the SME.
 }

\section{Discussion}
\label{Discussion}


The above Lagrangian is quite natural but obviously a rough toy model, proposed mostly to stimulate the curiosity of the  reader. Especially if he is interested in the idea of a Weyl/scale invariant early cosmology where the Weyl gauge symmetry is  subsequently spontaneously broken. This toy model shows that one can have a Weyl symmetry that is neither spontaneously broken nor gauge fixed, but \emph{erased} by an adequate redistribution of the d.o.f via dressing.  

In this respect, and as a comment on existing literature, we notice that our work supports the analysis of Jackiw and Pi \cite{Jackiw-Pi2015} who assert that the Weyl symmetry in recently introduced cosmological models, notably \cite{Bars-et-al2014}, is `fake'' and has no dynamical consequences. They supported their claim by showing that in these models the Noether current associated with the Weyl symmetry vanishes. But this same argument would lead to claim that the Weyl symmetry in Weyl/conformal gravity is fake as well: Indeed, almost simultaneously Campigotto and Fatibene have shown \cite{Campigotto-Fatibene2015} that  the Noether current also vanishes there. In our view  Jackiw and Pi  are  right, but the crucial argument is that in the models they discuss, a local dressing field - the dilaton - could be used to erase the Weyl symmetry, while there is no such dressing field in Weyl gravity. As they themselves remark at the end of their paper:
\begin{quote}
\emph{Introducing a spurion field and dressing up a model to appear gauge invariant is what we call a fake gauge invariance. [...] It will be interesting to find the symmetry current in a conventional Weyl invariant model, built on the square of the Weyl tensor. There the
symmetry is again local, but no scalar field is present to absorb the “gauge freedom.”}
\end{quote}

 What they call a \emph{fake} gauge symmetry is otherwise known as an \emph{artificial} gauge symmetry. The distinction artificial \emph{vs} substantial gauge symmetries is a very significant one that is mostly discussed within philosophy of physics \cite{Pitts2008, Pitts2009}, and is a direct extension of the long standing reflexion regarding the physical meaning of general covariance in General Relativity that was required to address the notorious Kretschmann objection. It happens that the DFM provides a simple criterion to detect artificial gauge symmetries: If a theory contains a local dressing field, then its gauge symmetry is artificial \cite{Francois2018}. 
In that regard, Jackiw and Pi correctly point out that in the models under study, the dilaton is a local dressing field (which they call a spurion field) so that the Weyl symmetry is artificial, or fake. Whereas in Weyl/conformal gravity there is no such local dressing field, so its Weyl symmetry is a \emph{substantial} gauge symmetry. Similar observations in the same period have been made by Hertzberg \cite{Hertzberg2015} and are likewise plainly interpreted within the framework of the DFM and the present work. More recent papers by `t~Hooft \cite{tHooft2015, tHooft2017}, where Weyl symmetry is supposed to be spontaneously broken, are also vulnerable to the same objection: The dilaton introduced is a local dressing field, the Weyl symmetry of the model is then artificial and is therefore of questionable physical meaning.


\section*{Acknowledgement}  

The author thanks Roberto Percacci (SISSA, Italy) for drawing his attention to the paper \cite{Jackiw-Pi2015} by Jackiw and Pi. This work was supporter by a postdoctoral grant of the Complexys Institut from the University of Mons, Belgium. 


{
\normalsize 
 \bibliography{Biblio1}

\begin{thebibliography}{10}

\bibitem{Hinterbichler&Khoury2012}
K.~Hinterbichler and J.~Khoury.
\newblock The pseudo-conformal universe: scale invariance from spontaneous
  breaking of conformal symmetry.
\newblock {\em Journal of Cosmology and Astroparticle Physics}, 2012(04):023,
  2012.

\bibitem{Bars-et-al2014}
I.~Bars, P.~Steinhardt, and N.~Turok.
\newblock Local conformal symmetry in physics and cosmology.
\newblock {\em Phys. Rev. D}, 89:043515, Feb 2014.

\bibitem{tHooft2017}
G.~t~Hooft.
\newblock Local conformal symmetry in black holes, standard model, and quantum
  gravity.
\newblock {\em International Journal of Modern Physics D}, 26(03):1730006,
  2017.

\bibitem{Atiyah_et_al2017}
M.~Atiyah, M.~Dunajski, and L.~Mason.
\newblock Twistor theory at fifty: from contour integrals to twistor strings.
\newblock {\em Proceedings of the Royal Society of London A: Mathematical,
  Physical and Engineering Sciences}, 473(2206), 2017.

\bibitem{Speziale2014}
S.~Speziale.
\newblock Loop quantum gravity, twistors, and some perspectives on the problem
  of time.
\newblock {\em EPJ Web of Conferences}, 71:00123, 2014.

\bibitem{Bailey-et-al94}
T.N. Bailey, M.G. Eastwood, and A.R. Gover.
\newblock Thomas's structure bundle for conformal, projective and related
  structures.
\newblock {\em Rocky Mountain J. Math.}, 24(4):1191--1217, 12 1994.

\bibitem{Gover-Shaukat-Waldron09}
Gover.~A. R., Shaukat. A, and Waldron. A.
\newblock Tractors, mass and weyl invariance.
\newblock {\em Nuclear Physics B}, 812(3):424--455, May 2009.

\bibitem{Sharpe}
R.~W. Sharpe.
\newblock {\em Differential Geometry: Cartan's Generalization of Klein's
  Erlangen Program}, volume 166 of {\em Graduate text in Mathematics}.
\newblock Springer, 1996.

\bibitem{Cap-Slovak09}
Andreas Cap and Jan Slovak.
\newblock {\em Parabolic Geometries I: Background and General Theory}, volume~1
  of {\em Mathematical Surveys and Monographs}.
\newblock American Mathematical Society, 2009.

\bibitem{Attard-Lazz2016}
J.~Attard and S.~Lazzarini.
\newblock {A note on Weyl invariance in gravity and the Wess-Zumino
  functional}.
\newblock {\em Nuclear Physics B}, 2016.

\bibitem{Jackiw-Pi2015}
R.~Jackiw and S.~Y. Pi.
\newblock Fake conformal symmetry in conformal cosmological models.
\newblock {\em Phys. Rev. D}, 91:067501, Mar 2015.

\bibitem{Attard_et_al2017}
J.~Attard, J.~Fran\c{c}ois, S.~Lazzarini, and T.~Masson.
\newblock {\em {Foundations of Mathematics and Physics one Century After
  Hilbert : New Perspectives}}, chapter {The dressing field method of gauge
  symmetry reduction, a review with examples}.
\newblock Springer, 2018.

\bibitem{Francois2018}
J.~Fran\c{c}ois.
\newblock {Artificial vs Substantial Gauge Symmetries: a Criterion and an
  Application to the Electroweak Model}.
\newblock {\em {Philosophy of Science}}, 2018.

\bibitem{Attard-Francois2016_I}
J.~Attard and J.~Fran\c{c}ois.
\newblock {Tractors and Twistors from conformal Cartan geometry: a gauge
  theoretic approach I. Tractors}.
\newblock {\em {arXiv, to appear in Adv. Theor. Math. Phys.}}, 2018.

\bibitem{Attard-Francois2016_II}
J.~Attard and J.~Fran\c{c}ois.
\newblock {Tractors and Twistors from conformal Cartan geometry: a gauge
  theoretic approach II. Twistors}.
\newblock {\em Class. Quantum Grav.}, 34(8), March 2017.

\bibitem{Pollock2010}
M.~D Pollock.
\newblock {On the Dirac Equation in Curved Spacetime}.
\newblock {\em {Acta Physica Polonica B}}, 41(8):1827--1846, 2010.

\bibitem{Bargmann1932}
V.~Bargmann.
\newblock Bemerkungen zur allgemein-relativistischen fassung der
  quantentheorie.
\newblock {\em Sitzungsber. Preuss. Akad. Wiss.}, 1932(XXIV):346--354, 1932.
\newblock As quoted by Pauli in \cite{Pauli1933-II} p.340 (in German). A
  textbook account is \cite{Sexl-Urbantke-2001} p.268. See also
  \cite{Pollock2010} or \cite{Arminjon-Reifler-2010}.

\bibitem{Francois2014}
J.~Fran\c{c}ois.
\newblock {\em {Reduction of gauge symmetries: a new geometrical approach}}.
\newblock Thesis, {Aix-Marseille Universit{\'e}}, September 2014.

\bibitem{collins_perez_sudarsky_2009}
J.~Collins, A.~Perez, and D.~Sudarsky.
\newblock {\em {Lorentz invariance violation and its role in Quantum Gravity
  phenomenology}}, pages 528--547.
\newblock Cambridge University Press, 2009.

\bibitem{Colladay-Kostelecky1998}
D.~Colladay and V.~Alan Kosteleck{\'y}.
\newblock Lorentz-violating extension of the standard model.
\newblock {\em Phys. Rev. D}, 58:116002, Oct 1998.

\bibitem{Mattingly2005}
D.~Mattingly.
\newblock {Modern Tests of Lorentz Invariance}.
\newblock {\em Living Reviews in Relativity}, 8(1):5, Sep 2005.

\bibitem{Greenberg2002}
O.~W. Greenberg.
\newblock {CPT Violation Implies Violation of Lorentz Invariance}.
\newblock {\em Phys. Rev. Lett.}, 89:231602, Nov 2002.

\bibitem{Bertolami-et-al1997}
O.~Bertolami, Don Colladay, V.Alan Kosteleck{\'y}, and R.~Potting.
\newblock {CPT violation and baryogenesis}.
\newblock {\em Physics Letters B}, 395(3):178--183, 1997.

\bibitem{Carroll-Shu2006}
S.~Carroll and J.~Shu.
\newblock {Models of baryogenesis via spontaneous Lorentz violation}.
\newblock {\em Phys. Rev. D}, 73:103515, May 2006.

\bibitem{Campigotto-Fatibene2015}
M.~Campigotto and L.~Fatibene.
\newblock Gauge natural formulation of conformal gravity.
\newblock {\em Annals of Physics}, 354:328 -- 337, 2015.

\bibitem{Pitts2008}
J.~B. Pitts.
\newblock {\em General Covariance, Artificial Gauge Freedom and Empirical
  Equivalence.}
\newblock PhD thesis, Graduate School of the University of Notre Dame, 2008.

\bibitem{Pitts2009}
J.~B. Pitts.
\newblock Empirical equivalence, artificial gauge freedom and a generalized
  kretschmann objection.
\newblock November 2009.

\bibitem{Hertzberg2015}
M.~P. Hertzberg.
\newblock Inflation, symmetry, and b-modes.
\newblock {\em {Physics Letters B}}, 745:118 -- 124, 2015.

\bibitem{tHooft2015}
G.~t~Hooft.
\newblock Local conformal symmetry: The missing symmetry component for space
  and time.
\newblock {\em International Journal of Modern Physics D}, 24(12):1543001,
  2015.

\bibitem{Pauli1933-II}
W.~Pauli.
\newblock {{\"U}ber die Formulierung der Naturgesetze mit f{\"u}nf homogenen
  Koordinaten. Teil II: Die Diracschen Gleichungen f{\"u}r die Materiewellen}.
\newblock {\em Annalen der Physik}, 18:337--372, 1933.

\bibitem{Sexl-Urbantke-2001}
R.~U. Sexl and H.~K. Urbantke.
\newblock {\em {Relativity, Groups, Particles: Special Relativity and
  Relativistic Symmetry in Field and Particle Physics}}.
\newblock {Springer-Verlag Wien}, 2001.

\bibitem{Arminjon-Reifler-2010}
Mayeul Arminjon and Frank Reifler.
\newblock {Basic quantum mechanics for three Dirac equations in a curved
  spacetime}.
\newblock {\em {Brazilian Journal of Physics}}, 40:242 -- 255, 06 2010.

\end{thebibliography}
}

\end{document}